\documentclass[aps,reprint,amsmath,amssymb,nofootinbib,floatfix,showpacs]{revtex4-1}

\usepackage[colorlinks,citecolor=blue,linkcolor=blue]{hyperref}
\usepackage{graphicx}
\usepackage{dcolumn}
\usepackage{bm}
\usepackage{mathtools}
\usepackage{txfonts}

\begin{document}

\title{Role of nonlinear anisotropic damping in the magnetization dynamics of topological solitons}

\author{Joo-Von Kim}
\email{joo-von.kim@u-psud.fr}
\affiliation{Institut d'Electronique Fondamentale, Univ. Paris-Sud, 91405 Orsay, France}
\affiliation{CNRS, UMR 8622, 91405 Orsay, France}

\date{\today}

\begin{abstract}
The consequences of nonlinear anisotropic damping, driven by the presence of Rashba spin-orbit coupling in thin ferromagnetic metals, are examined for the dynamics of topological magnetic solitons such as domain walls, vortices, and skyrmions. The damping is found to affect Bloch and N{\'e}el walls differently in the steady state regime below Walker breakdown and leads to a monotonic increase in the wall velocity above this transition for large values of the Rashba coefficient. For vortices and skyrmions, a generalization of the damping tensor within the Thiele formalism is presented. It is found that chiral components of the damping affect vortex- and hedgehog-like skyrmions in different ways, but the dominant effect is an overall increase in the viscous-like damping.
\end{abstract}


\maketitle

\section{Introduction}
Dissipation in magnetization dynamics is a longstanding problem in magnetism~\cite{Landau:1935tr, Sparks:1964vb, Gilbert:2004gx}. For strong ferromagnets such as cobalt, iron, nickel, and their alloys, a widely used theoretical approach to describe damping involves a local viscous form due to Gilbert for the Landau-Lifshitz equation of motion,
\begin{equation}
\frac{\partial \mathbf{m}}{\partial t} = -\gamma_0 \mathbf{m} \times \mathbf{H}_{\rm eff} + \alpha_0 \mathbf{m} \times \frac{\partial \mathbf{m}}{\partial t},
\label{eq:LLG}
\end{equation}
which appears as the second term on the right hand side, proportional to the damping constant $\alpha_0$.  This equation describes the damped magnetization precession about a local effective field $\mathbf{H}_{\rm eff} = - (1/\mu_0 M_s) \delta U / \delta \mathbf{m} $, which is given by a variational derivative of the magnetic energy $U$ with respect to the magnetization field described by the unit vector $\mathbf{m}$, with $\gamma_0 = \mu_0 \gamma$ being the gyromagnetic constant and $M_s$ is the saturation magnetization.  Despite the multitude of physical processes that underlie dissipation in such materials, such as the scattering of magnons with electrons, phonons, and other magnons, the form in Eq.~(\ref{eq:LLG}) has proven to be remarkably useful for describing a wide range of dynamical phenomena from ferromagnetic resonance to domain wall motion.

One feature of the dissipative dynamics described in Eq.~(\ref{eq:LLG}) is that it is local, i.e., the damping torque only depends on the local magnetization and its time dependence. With the advent of magnetic heterostructures, however, this restriction of locality has been shown to be inadequate for systems such as metallic multilayers in which nonlocal processes can be important~\cite{Tserkovnyak:2005fr}. A striking example involves spin pumping, which describes how spin angular momentum can be dissipated in adjacent magnetic or normal metal layers through the absorption of spin currents generated by a precessing magnetization~\cite{Tserkovnyak:2002ju, Mills:2003hg}.  Early experimental observations of this phenomena involved iron films sandwiched by silver layers~\cite{Hurdequint:1991vj} and permalloy films in close proximity with strong spin-orbit normal metals such as palladium and platinum~\cite{Mizukami:2001fy, Mizukami:2002hh}, where ferromagnetic resonance line widths were shown to depend strong on the composition and thickness of the adjacent layers. Such observations also spurred other studies involving ferromagnetic multilayers separated by normal metal spacers, where spin pumping effects can lead to a dynamic coupling between the magnetization in different layers~\cite{Hurdequint:1991tt, Urban:2001hr}. In the context of damping, such dynamic coupling was shown to give rise to a configuration dependent damping in spin-valve structures~\cite{Kim:2005gr, Joyeux:2011hd}.

A generalization of the spin-pumping picture in the context of dissipation was given by Zhang and Zhang, who proposed that spin currents generated within the ferromagnetic material itself can lead to an additional contribution to the damping, provided that large magnetization gradients are present~\cite{Zhang:2009ig}. This theory is based on an $sd$ model in which the local moments (4$d$) are exchange coupled to the delocalized conduction electrons (3$s$), which are treated as a free electron gas.  The spin current ``pumped'' at one point in the material by the precessing local moments are dissipated at another if the current encounters strong spatial variations in the magnetization such as domain walls or vortices -- a mechanism that can be thought of as the reciprocal process of current-induced spin  torques in magnetic textures~\cite{Berger:1984ti, Zhang:2004hs, ClaudioGonzalez:2012er, Manchon:2011vl}. For this reason, the mechanism is referred to as ``feedback'' damping since the pumped spin currents generated feed back into the magnetization dynamics in the form of a dissipative torque. This additional contribution is predicted to be both nonlinear and nonlocal, and can have profound consequences for the dynamics of topological solitons such as domain walls and vortices as a result of the spatial gradients involved. Indeed, recent experiments on vortex wall motion in permalloy stripes indicate that such nonlinear contributions can be significant and be of the same order of magnitude as the usual Gilbert damping characterized by $\alpha_0$ in Eq.~(\ref{eq:LLG})~\cite{Weindler:2014et}.

An extension to this feedback damping idea was proposed recently by Kim and coworkers, who considered spin pumping involving a conduction electron system with a Rashba spin-orbit coupling (RSOC)~\cite{Kim:2012kx}. By building upon the Zhang-Zhang formalism, it was shown that the feedback damping can be expressed as a generalization of the Landau-Lifshitz equation~\cite{Zhang:2009ig, Kim:2012kx},
\begin{equation}
\frac{\partial \mathbf{m}}{\partial t} = -\gamma_0 \mathbf{m} \times \mathbf{H}_{\rm eff} + \mathbf{m} \times \mathcal{D}_{\rm LL}(\mathbf{m}) \cdot \frac{\partial \mathbf{m}}{\partial t},
\end{equation}
where the $3 \times 3$ matrix $\mathcal{D}_{\rm LL}$ represents the generalized damping tensor, which can be expressed as~\cite{Kim:2012kx}
\begin{equation}
\mathcal{D}_{\textrm{LL}}^{ij} = \alpha_0 \delta_{ij} + \eta \sum_{k} \left( F_{ki} + \tilde{\alpha}_{\rm R} \epsilon_{3ki}\right)\left( F_{kj} + \tilde{\alpha}_{\rm R} \epsilon_{3kj}\right).
\label{eq:nad}
\end{equation}
Here, $\alpha_0$ is the usual Gilbert damping constant, $\eta = g \mu_B \hbar G_0 / (4 e^2 M_s)$ is a constant related to the conductivity $G_0$ of the spin bands~\cite{Zhang:2009ig}, $F_{ki} = \left( \partial \mathbf{m} / \partial x_k \right)_i$ are components of the spatial magnetization gradient, $\tilde{\alpha}_{\rm R} = 2 \alpha_{\rm R} m_e / \hbar^2$ is the scaled Rashba coefficient, $\epsilon_{ijk}$ is the Levi-Civita symbol, and the indices $(ijk)$ represent the components $(xyz)$ in Cartesian coordinates. In addition to the nonlinearity present in the Zhang-Zhang picture, the inclusion of the $\alpha_{\rm R}$ term results in an anisotropic contribution that is related to the underlying symmetry of the Rashba interaction. Numerical estimates based on realistic parameters suggest that the Rashba contribution can be much larger than the nonlinear contribution $\eta$ alone~\cite{Kim:2012kx}, which may have wide implications for soliton dynamics in ultrathin ferromagnetic films with perpendicular magnetic anisotropy, such as Pt/Co material systems, in which large spin-orbit effects are known to be present.

In this article, we explore theoretically the consequences of the nonlinear anisotropic damping given in Eq.~(\ref{eq:nad}) on the dynamics of topological magnetic solitons, namely domain walls, vortices, and skyrmions, in which spatial  gradients can involve 180$^\circ$ rotation of the magnetization vector over length scales of 10 nm. In particular, we examine the role of chirality in the Rashba-induced contributions to the damping, which are found to affect chiral solitons in different ways. This article is organized as follows. In Section~\ref{sec:dw}, we discuss the effects of nonlinear anisotropic damping on the dynamics of Bloch and N{\'e}el domain walls, where the latter is stabilized by the Dzyaloshinskii-Moriya interaction. In Section~\ref{sec:vorsky}, we examine the consequences of this damping for vortices and skyrmions, and we derive a generalization to the damping dyadic appearing in the Thiele equation of motion. Finally, we present some discussion and concluding remarks in Section~\ref{sec:discussion}.

\section{\label{sec:dw}Bloch and N{\'e}el domain walls}
The focus of this section are domain walls in ultrathin films with perpendicular magnetic anisotropy. Consider a 180$^\circ$ domain wall representing a boundary separating two oppositely magnetized domains along the $x$ axis, with $z$ being the uniaxial anisotropy axis that is perpendicular to the film plane. We assume that the magnetization remains uniform along the $y$ axis. The unit magnetization vector $\mathbf{m}(x,t)$ can be parametrized in spherical coordinates $(\theta, \phi)$, such that $\mathbf{m} = (\sin\theta \cos\phi, \sin\theta \sin\phi, \cos\theta)$. With this definition, the spherical angles for the domain wall profile can be written as
\begin{align}
	\theta(x,t) &= 2 \tan^{-1}\exp\left(\pm \frac{x-X_0(t)}{\Delta} \right), \nonumber \\
	\phi(x,t) &= \phi_0(t),
	\label{eq:dw}
\end{align}
where $X_0(t)$ denotes the position of the domain wall, $\Delta = \sqrt{A/K_0}$ represents the wall width parameter that depends on the exchange constant $A$ and the effective uniaxial anisotropy $K_0$, and the azimuthal angle $\phi_0(t)$ is a dynamic variable but spatially uniform. The anisotropy constant, $K_0 = K_u - \mu_0 M_s^2/2$,  involves the difference between the magnetocrystalline ($K_u$) and shape anisotropies relevant for an ultrathin film. In this coordinate system, a static Bloch wall is given by $\phi_0 = \pm \pi/2$, while a static N{\'e}el wall is given by $\phi_0 = 0,\pi$. A positive sign in the argument of the exponential function for $\theta$ in Eq.~(\ref{eq:dw}) describes an up-to-down domain wall profile going along the $+x$ direction, while a negative sign represents a down-to-up wall.

To determine the role of the nonlinear anisotropic damping term in Eq.~(\ref{eq:nad}) on the wall dynamics, it is convenient to compute the dissipation function $W(\dot{X}_0,\dot{\phi}_0)$ for the wall variables, where the notation $\dot{X}_0 \equiv \partial_t X_0$, etc., denotes a time derivative. The dissipation function per unit surface area is given by
\begin{equation}
W \left( \dot{X_0},\dot{\phi_0} \right) = \frac{M_s}{2\gamma} \int_{-\infty}^{\infty} dx \; \dot{m}_i  \, \mathcal{D}_{\rm LL}^{ij}(\mathbf{m}) \, \dot{m}_j, 
\label{eq:dissfn}
\end{equation}
where $m_i = m_i\left[x-X_0(t),\phi_0(t)\right]$ and the Einstein summation convention is assumed. By using the domain wall ansatz (\ref{eq:dw}), the integral in Eq.~(\ref{eq:dissfn}) can be evaluated exactly to give $W = W_0 + W_{\rm NL}$, where $W_0$ represents the usual (linear) Gilbert damping,
\begin{equation}
W_0 = \alpha_0 \frac{M_s \Delta}{\gamma} \left( \frac{\dot{X}_0^2 }{\Delta^2} + \dot{\phi}^2_0  \right),
\end{equation}
while $W_{\rm NL}$ is the additional contribution from the nonlinear anisotropic damping,
\begin{equation}
\begin{split}
W_{\rm NL} &= \frac{M_s \Delta}{\gamma} \left[ \frac{1}{3}\alpha_3 \sin^2\phi_0(t) \frac{\dot{X}_0^2 }{\Delta^2}  \right. \\
&+ \left.  \left(  \frac{2}{3} \alpha_1  \pm \frac{\pi}{2} \alpha_2  \cos\phi_0(t) + \alpha_3 \cos^2\phi_0(t)  \right)\dot{\phi}^2_0 \right],
\end{split}
\end{equation}
where $\alpha_1 \equiv \eta/\Delta^2$, $\alpha_2 \equiv \eta \, \tilde{\alpha}_{\rm R}/\Delta$, and $\alpha_3 \equiv \eta \, \tilde{\alpha}_{\rm R}^2$ are dimensionless nonlinear damping constants. In contrast to the linear case, the nonlinear anisotropic dissipation function exhibits a configuration-dependent dissipation rate where the prefactors of the $\dot{X}_0^2$ and $\dot{\phi}_0^2$ terms depend explicitly on $\phi_0(t)$.

In addition to the nonlinearity a \emph{chiral} damping term, proportional to $\alpha_2$, appears as a result of the Rashba interaction and is linear in the Rashba coefficient $\alpha_{\rm R}$. The sign of this term depends on the sign chosen for the polar angle $\theta$ in the wall profile (\ref{eq:dw}). To illustrate the chiral nature of this term, we consider small fluctuations about the static configuration by writing $\phi_0(t) = \phi_0 + \delta \phi(t)$, where $\delta \phi(t) \ll \pi$ is a small angle. This approximation is useful for the steady state regime below Walker breakdown. For up-to-down Bloch walls ($\phi_0 = \pm \pi/2$), the nonlinear part of the dissipation function to first order in $\delta \phi(t)$ becomes
\begin{equation}
W_{\rm NL,Bloch} \approx \frac{M_s \Delta}{\gamma} \left[ \frac{\alpha_3}{3} \frac{\dot{X}_0^2 }{\Delta^2}  + \left( \frac{2 \alpha_1}{3} + C_x\frac{\pi \alpha_2}{2}  \, \delta \phi(t) \right)  \, \dot{\phi}^2_0  \right].
\label{eq:dissfn_Bloch_lin}
\end{equation}
The quantity $C_i = \pm 1$ is a component of the chirality vector~\cite{Braun:2012kw},
\begin{equation}
\mathbf{C} = \frac{1}{\pi} \int_{-\infty}^{\infty} dx \, \mathbf{m} \times \partial_x \mathbf{m},
\end{equation}
which characterizes the handedness of the domain wall. For a right-handed Bloch wall, $\phi_0 = -\pi/2$ and the only nonvanishing component is $C_x = 1$, while for a left-handed wall ($\phi_0 = -\pi/2$) the corresponding value is  $C_x = -1$. Thus, the term proportional to $\alpha_2$ depends explicitly on the wall chirality. Similarly for up-to-down N{\'e}el walls, the same linearization about the static wall profile leads to 
\begin{equation}
W_{\rm NL,Neel} \approx \frac{M_s \Delta}{\gamma} \left( \frac{2 \alpha_1}{3} + C_y\frac{\pi \alpha_2}{2}  + \alpha_3 \right)  \dot{\phi}^2_0,
\end{equation}
where $C_y = 1$ for a right-handed N{\'e}el wall ($\phi_0 = 0$) and $C_y = -1$ for its left-handed counterpart ($\phi_0 = \pi$). Since the fluctuation $\delta \phi(t)$ is taken to be small, the chiral damping term is more pronounced for N{\'e}el walls in the steady-state velocity regime since it does not depend on the fluctuation amplitude $\delta \phi(t)$ as in the case of Bloch walls.

To better appreciate the magnitude of the chirality-dependent damping term, it is instructive to estimate numerically the relative magnitudes of the nonlinear damping constants $\alpha_1, \alpha_2, \alpha_3$. Following [Ref.~\onlinecite{Kim:2012kx}], we assume $\eta = 0.2$ nm$^2$ and $\alpha_{\rm R}  = 10^{-10}$ eV m. If we suppose $\Delta = 10$ nm, which is consistent with anisotropy values measured in ultrathin films with perpendicular anisotropy~\cite{Burrowes:2013jl},  the damping constants can be evaluated to be $\alpha_1 = 0.002$, $\alpha_2 = 0.052$, and $\alpha_3 = 1.37$. Since $\alpha_0$ varies between 0.01--0.02~\cite{Devolder:2013hq} and 0.1--0.3~\cite{Schellekens:2013it} depending on the material system, the chiral term $\alpha_2$ is comparable to Gilbert damping in magnitude, but remains almost an order of magnitude smaller than the nonlinear component $\alpha_3$ that provides the dominant contribution to the overall damping.

The full equations of motion for the domain wall dynamics can be obtained using a Lagrangian formalism that accounts for the dissipation given by $W$~\cite{Thiaville:2002gy, LeMaho:2009ee}. For the sake of simplicity, we will focus on wall motion driven by magnetic fields alone, where a spatially-uniform magnetic field $H_z$ is applied along the $+z$ direction. In addition, we include the Dzyaloshinskii-Moriya interaction appropriate for the geometry considered~\cite{Bogdanov:2001hr, Thiaville:2012ia} when considering the dynamics of N{\'e}el walls. From the Euler-Lagrange equations with the Rayleigh dissipation function,
\begin{equation}
\frac{d}{dt} \frac{\partial L}{\partial \dot{X}_0} - \frac{\partial L}{\partial X_0} + \frac{\partial W}{\partial \dot{X}_0} = 0,
\end{equation}
with an analogous expression for $\phi_0$, the equations of motion for the wall coordinates are found to be
\begin{equation}
\dot{\phi}_0 + \left( \alpha_0+ \frac{\alpha_3}{3} \sin^2\phi_0 \right)\frac{\dot{X}_0}{\Delta}  = \gamma_0 H_z,
\label{eq:phidot}
\end{equation}
\begin{multline}
\frac{\dot{X}_0}{\Delta} - \left( \alpha_0 + \frac{2\alpha_1}{3} + \frac{\pi \alpha_2}{2} \cos\phi_0 + \alpha_3 \cos^2\phi_0  \right)\dot{\phi}_0 \\
=-\gamma_0 \left( \frac{\pi}{2} \frac{D_{\rm ex}}{\mu_0 M_s \Delta}  + \frac{2K_\perp}{\mu_0 M_s} \cos\phi_0  \right) \sin\phi_0,
\label{eq:qdot}
\end{multline}
where $D_{\rm ex}$ is the Dzyaloshinskii-Moriya constant~\cite{Thiaville:2012ia} and $K_\perp$ represents a hard-axis anisotropy that results from volume dipolar charges. The Dzyaloshinskii-Moriya interaction (DMI) is present in ultrathin films in contact with a strong spin-orbit coupling material~\cite{Fert:1980hr, Fert:1990} and favors a N{\'e}el-type wall profile~\cite{Bode:2007em, Heide:2008da}. The DMI itself can appear as a consequence of the Rashba interaction and therefore its inclusion here is consistent with the nonlinear anisotropic damping terms used~\cite{Imamura:2004hn, Kim:2012kx, Kim:2013gm}.

\begin{figure}
\centering\includegraphics[width=7.0cm]{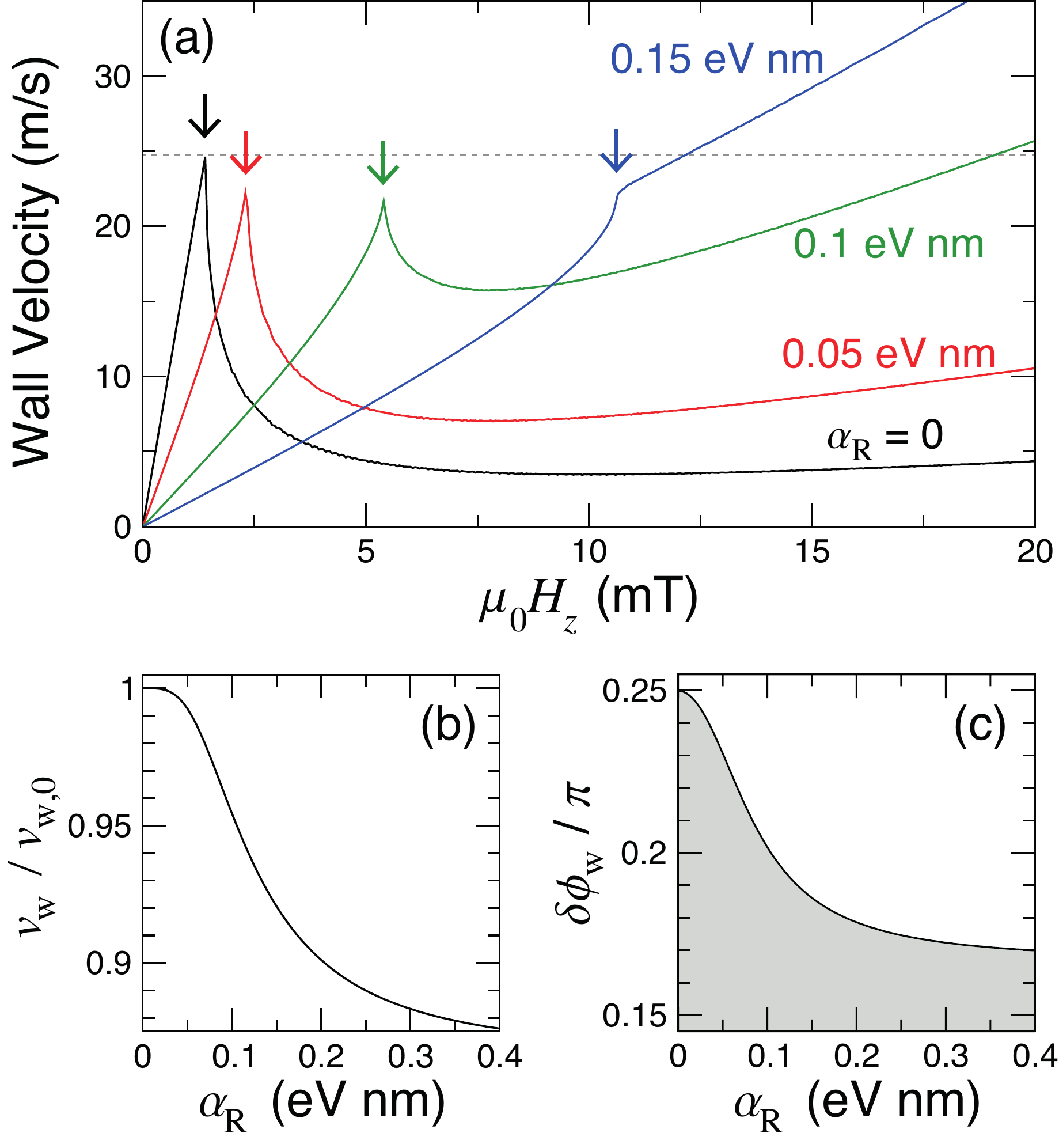}
\caption{(Color online) Bloch wall dynamics. (a) Steady-state domain wall velocity, $\langle \dot{X}_0 \rangle$, as a function of perpendicular applied magnetic field, $\mu_0 H_z$, for several values of the Rashba coefficient, $\alpha_{\rm R}$. The horizontal dashed line indicates the Walker velocity and the arrows indicate the Walker transition. (b) The ratio between the Walker velocity, $v_{\rm W}$, to its linear damping value, $v_{\rm W,0}$, as a function of $\alpha_{\rm R}$. (c) Deviation in the wall angle from rest at the Walker velocity, $\delta \phi_{\rm W}$, as a function of $\alpha_R$}
\label{fig:BlochWall}
\end{figure}

\begin{figure}
\centering\includegraphics[width=7.0cm]{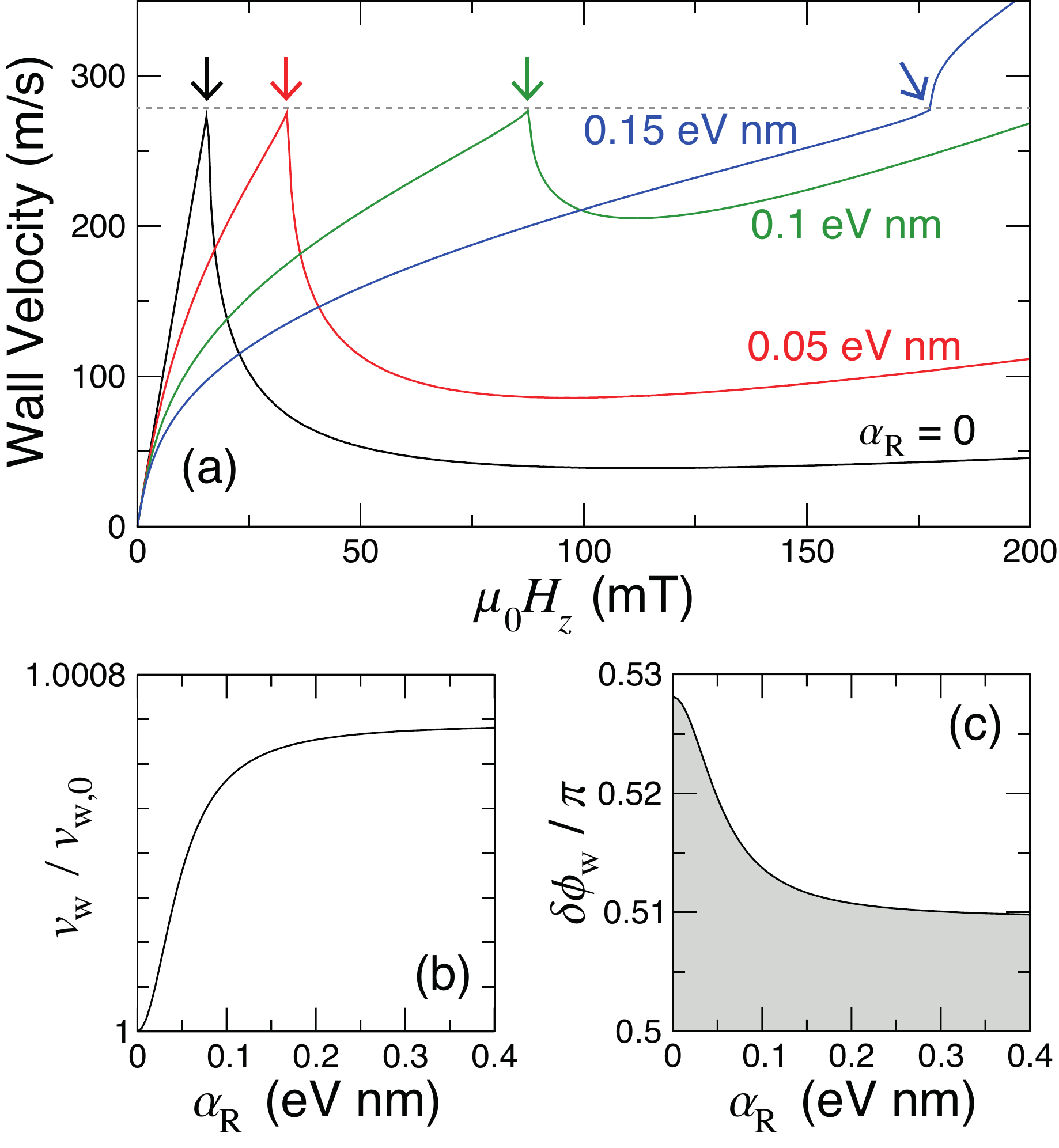}
\caption{(Color online) Dzyaloshinskii (N{\'e}el) wall dynamics. (a) Steady-state domain wall velocity, $\langle \dot{X}_0 \rangle$, as a function of perpendicular applied magnetic field, $\mu_0 H_z$, for several values of the Rashba coefficient, $\alpha_{\rm R}$. The horizontal dashed line indicates the Walker velocity and the arrows indicate the Walker transition. (b) The ratio between the Walker velocity, $v_{\rm W}$, to its linear damping value, $v_{\rm W,0}$, as a function of $\alpha_R$. (c) The wall angle at the Walker velocity, $\phi_{\rm W}$, as a function of $\alpha_R$}
\label{fig:NeelWall}
\end{figure}

Results from numerical integration of these equations of motion for Bloch and N{\'e}el walls are presented in Figs.~\ref{fig:BlochWall} and \ref{fig:NeelWall}.
We used parameters consistent with ultrathin films with perpendicular anisotropy, namely $\alpha_0 = 0.1$, $M_s =$ 1 MA/m, $\Delta =$ 10 nm, and $K_\perp = \mu_0 N_x M_s^2 / 2$ with the demagnetization factor $N_x = 0.02$~\cite{Thiaville:2012ia}. To study the dynamics of the Dzyaloshinskii (N{\'e}el) wall we assumed a value of $D_{\rm ex} = 1$ mJ/m$^2$, which is much stronger than the volume dipolar interaction represented by $K_\perp$ and is of the same order of magnitude as values determined by Brillouin light spectroscopy in Pt/Co/Al$_{2}$O$_{3}$ films~\cite{Belmeguenai:2015ui}. As in the discussion on numerical estimates above, we assumed $\eta = 0.2$ nm$^2$ but considered several different values for the Rashba coefficient $\alpha_{\rm R}$. The steady-state domain wall velocity, $\langle \dot{X}_0 \rangle$, was computed as a function of the perpendicular applied magnetic field, $H_z$. In the precessional regime above Walker breakdown in which $\phi_{0}(t)$ becomes a periodic function in time, $\langle \dot{X}_0 \rangle$ is computed by averaging the wall displacement over few hundred periods of precession.

For the Bloch case [Fig.~\ref{fig:BlochWall}(a)], the Walker field is observed to increase with the Rashba coefficient, which is consistent with the overall increase in damping experienced by the domain wall. However, there are two features that differ qualitatively from the behavior with linear damping. First, the Walker velocity is not attained for finite $\alpha_{\rm R}$, where the peak velocity at the Walker transition is below the value reached for $\alpha_{\rm R} = 0$.  This is shown in more detail in Fig.~\ref{fig:BlochWall}(b), where the ratio between the Walker velocity, $v_{\rm W}$, and its linear damping value, $v_{\textrm{W},0}$, is shown as a function of $\alpha_{\rm R}$. The Walker limit is set by the largest extent to which the wall angle $\phi_0$ can deviate from its equilibrium value, $\phi_{0,\textrm{eq}}$.  By assuming $\dot{\phi} = 0$ in the linear regime, we can determine this limit by rearranging Eqs.~\ref{eq:phidot} and \ref{eq:qdot} to obtain the following relationship for the Bloch wall,
\begin{equation}
\frac{2 H_z}{N_x M_s} = -\left( \alpha_0 + \frac{\alpha_3}{3} \sin^2\phi_0  \right) \sin 2\phi_0.
\end{equation}
The angle $\phi_0 = \phi_{\rm W}$ for which the right hand side of this equation is an extremum determines the Walker limit. In Fig.~\ref{fig:BlochWall}(c), we present this limit in terms of the deviation angle, $\delta \phi_{\rm W} \equiv | \phi_{\rm W} - \phi_{0,\textrm{eq}} |$, which is shown as a function of $\alpha_{\rm R}$. As the Rashba parameter is increased, the maximum wall tilt possible in the linear regime decreases from the linear damping value of $\pi/4$, which results in an overall reduction in the Walker velocity. Second, the field dependence of the wall velocity below Walker breakdown is nonlinear and exhibits a slight convex curvature, which becomes more pronounced as $\alpha_{\rm R}$ increases. This curvature can be understood by examining the wall mobility under fields, which can be deduced from Eq. (\ref{eq:phidot}) by setting $\dot{\phi}=0$,
\begin{equation}
\dot{X}_0 = \frac{\gamma_0 \Delta}{\alpha_0 + \left(\alpha_3/3\right) \sin^2{\phi_0}} H_z.
\label{eq:mobility}
\end{equation}
Since the angle $\phi_0$ for Bloch walls varies from its rest value of $\phi_{0,\textrm{eq}} = \pm \pi/2$ at zero field to $\phi_{\rm W}$ at the Walker field, the $\sin^2\phi_0$ term in the denominator decreases from its maximum value of $\sin^2\phi_{0,\textrm{eq}} = 1$ at rest with increasing applied field and therefore an increase in the mobility is seen as $H_z$ increases, resulting in the convex shape of the velocity versus field relation below Walker breakdown.

It is interesting to note that the nonlinear damping terms affect the Dzyaloshinskii (N{\'e}el) wall motion differently. In contrast to the Bloch case, the Walker velocity for increasing $\alpha_{\rm R}$ slightly exceeds the linear damping value, which can be seen by the arrows marking the Walker transition in Fig.~\ref{fig:NeelWall}(a) and in detail in Fig.~\ref{fig:NeelWall}(b).
In addition, the field dependence of the velocity exhibits a concave curvature below breakdown, which can also be understood from Eq. (\ref{eq:mobility}) by considering that $\phi_0$ instead deviates from the rest value of $\phi_{0,\textrm{eq}} = 0$ or $\pi$ at zero field. As for the Bloch wall case, the deviation angle at breakdown is determined by the value of $\phi_0$ that gives an extremum for the right hand side of 
\begin{equation}
\frac{2 H_z}{N_x M_s} = -\left( \alpha_0 + \frac{\alpha_3}{3} \sin^2\phi_0  \right) \left( \frac{\pi D_{\rm ex}}{2 K_{\perp} \Delta} \cos\phi_0 +  \sin 2\phi_0 \right),
\end{equation}
and is also seen to decrease with increasing Rashba coefficient [Fig.~\ref{fig:NeelWall}(c)]. In contrast to the Bloch wall case, however, changes in $\phi_{\rm W}$ have a comparatively modest effect on the Walker velocity. The shape of the velocity versus field curve is consistent with experimental reports of field-driven domain wall motion in the Pt/Co (0.6 nm)/Al$_{2}$O$_3$ system~\cite{Miron:2011fn}, which possess a large DMI value~\cite{Belmeguenai:2015ui} and harbors N{\'e}el-type domain wall profiles at equilibrium~\cite{Tetienne:2015ka}.

As the preceding discussion shows, the differences in the field dependence of the wall velocity for the two profiles are a result of the DMI, rather than the chiral damping term that is proportional to $\alpha_2$. This was verified by setting $\alpha_2 = 0$ for the N{\'e}el wall case with $D \neq 0$, which did not modify the overall behavior of the field dependence of the velocity. In the one-dimensional approximation for the wall dynamics, the DMI enters the equations of motion like an effective magnetic field along the $x$ axis, which stabilizes the wall structure by minimizing deviations in the wall angle $\phi_0(t)$.

\section{\label{sec:vorsky}Vortices and skyrmions}
The focus of this section is on the dissipative dynamics of two-dimensional topological solitons such as vortices and skyrmions. The equilibrium magnetization profile for these micromagnetic objects are described by a nonlinear differential equation similar to the sine-Gordon equation, where the dispersive exchange interaction is compensated by dipolar interactions for vortices~\cite{Feldtkeller:1965uz, Gaididei:2010ez} and an additional uniaxial anisotropy for skyrmions~\cite{Kiselev:2011cm}.  The topology of vortices and skyrmions can be characterized by the skyrmion winding number $Q$,
\begin{equation}
Q = \frac{1}{4 \pi} \iint dx dy \; \mathbf{m} \cdot \left( \partial_x \mathbf{m} \times \partial_y \mathbf{m}  \right).
\end{equation}
While the skyrmion number for vortices ($Q = \pm 1/2$) and skyrmions ($Q = \pm 1$) are different, their dynamics are  qualitatively similar and can be described using the same formalism. For this reason, vortices and skyrmions will be treated on equal footing in what follows and distinctions between the two will only be drawn on the numerical values of the damping parameters.

\begin{figure}
\centering\includegraphics[width=8.5cm]{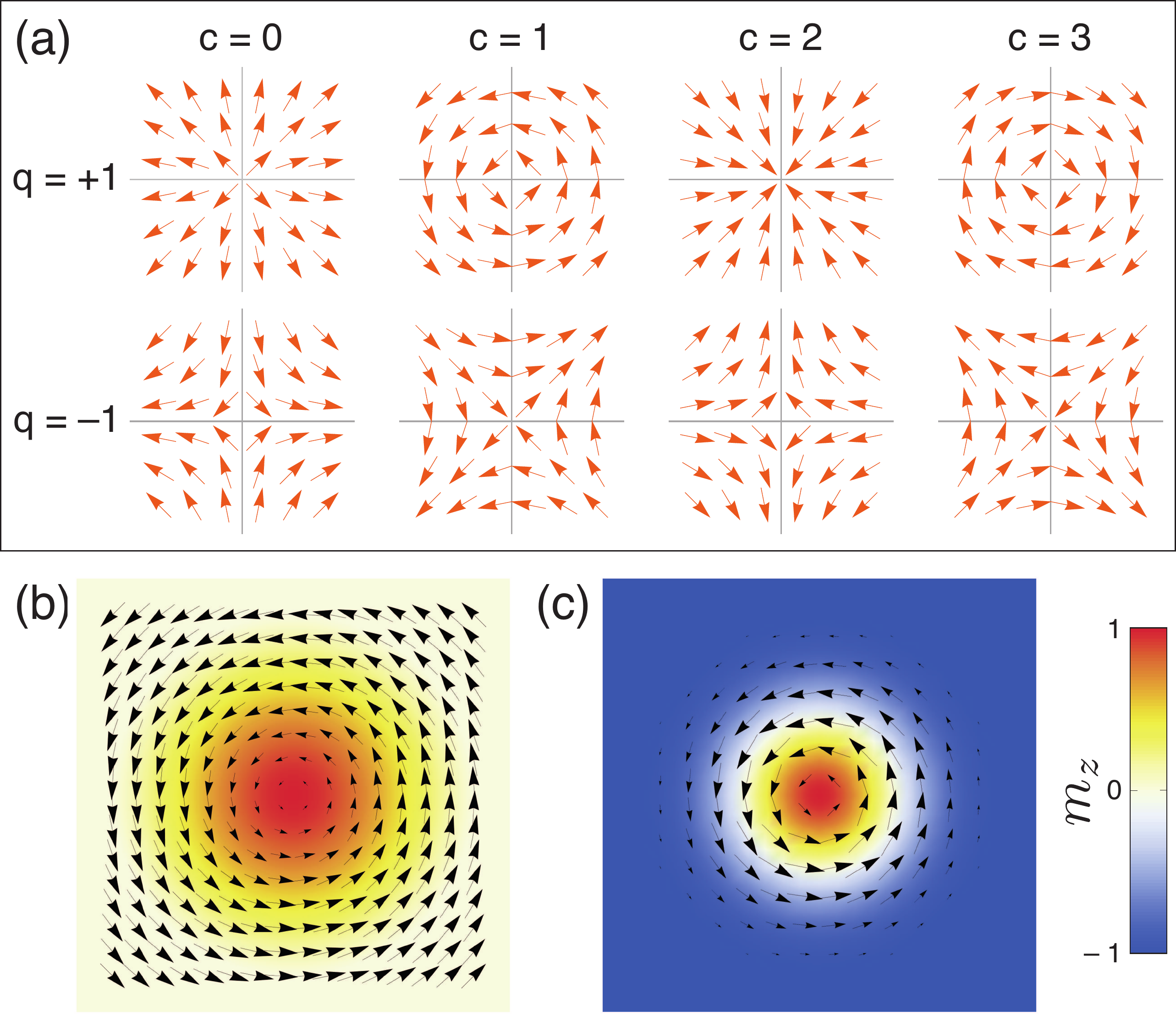}
\caption{(Color online) In-plane magnetization fields for vortices and skyrmions. (a) Vector fields given by $\phi(x,y)$ in (\ref{eq:vorsky}) for different values of $q$ and $c$. (b) Vortex  and (c) skyrmion for spin structure with $c=1, q=1$, where the arrows indicate the in-plane components ($m_{x,y}$) and the color code gives the perpendicular component of the magnetization ($m_z$). }
\label{fig:vortex_config}
\end{figure}

A key approximation used for describing vortex or skyrmion dynamics is the rigid core assumption, where it is assumed that the spin structure of the soliton remains unperturbed from its equilibrium state during motion.  Within this approximation, the dynamics is given entirely by the position of the core in the film plane, $\mathbf{X}_0(t) = \left[X_0(t),Y_0(t) \right]$, which allows the unit magnetization vector to be parametrized as
\begin{align}
	\theta(x,y,t) &= \theta_0\left[  \| \mathbf{x} - \mathbf{X}_0(t)  \|  \right], \nonumber \\
	\phi(x,y,t) &= q \tan^{-1}\left[\frac{y-Y_0(t)}{x-X_0(t)}\right] + c \frac{\pi}{2},
	\label{eq:vorsky}
\end{align}
where $q$ is a topological charge and $c$ is the chirality. An illustration of the magnetization field given by the azimuthal angle $\phi(x,y)$ is presented in Fig.~\ref{fig:vortex_config}. 
$q=1$ corresponds to a vortex or skyrmion, while $q = -1$ represents the antivortex or antiskyrmion.

The dynamics of a vortex or skyrmion in the rigid core approximation is given by the Thiele equation,
\begin{equation}
\mathbf{G} \times \dot{\mathbf{X}}_0 + \mathcal{D}_{\rm T} \cdot \dot{\mathbf{X}}_0 = -\frac{\partial U}{\partial \mathbf{X}_0},
\label{eq:Thiele}
\end{equation}
where
\begin{equation}
\mathbf{G} = \frac{M_s d}{\gamma} \iint dx dy \, \sin(\theta) \left( \nabla\theta \times \nabla\phi  \right)
\end{equation}
is the gyrovector and $U(\mathbf{X}_0)$ is the effective potential that is obtained from the magnetic Hamiltonian by integrating out the spatial dependence of the magnetization. The damping dyadic in the Thiele equation, $\mathcal{D}_{\rm T}$, can be obtained from the dissipation function in the rigid core approximation, $W(\dot{\mathbf{X}}_0)$, which is defined in the same way as in Eq.~(\ref{eq:dissfn}) but with the ansatz given in Eq.~(\ref{eq:vorsky}). For this system, it is more convenient to evaluate the dyadic by performing the integration over all space after taking derivatives with respect to the core velocity. In other words, the dyadic can be obtained using the Lagrangian formulation by recognizing that
\begin{equation}
\mathcal{D}_{\rm T} \cdot \dot{\mathbf{X}}_0 = \frac{M_s d}{2 \gamma}  \iint dx dy \; \frac{\partial}{\partial \dot{\mathbf{X}}_0} \left(\dot{m}_i  \, \mathcal{D}_{\rm LL}^{ij}(\mathbf{m}) \, \dot{m}_j \right).
\end{equation}
By using polar coordinates for the spatial coordinates, $(x,y)=(r \cos\varphi, r \sin\varphi)$, assuming translational invariance in the film plane, and integrating over $\varphi$, the damping dyadic is found to be
\begin{equation}
\mathcal{D}_{\rm T} = \frac{M_s d}{\gamma}  \left(  \left( \alpha_0 D_0 + \alpha_1 D_1 + \alpha_3 D_3 \right) \mathcal{I}  + \alpha_2 D_2  \left[\begin{array}{cc}a_{11} & 0 \\0 & a_{22}\end{array}\right] \right),
\label{eq:dyadic}
\end{equation}
where $\mathcal{I}$ is the $2\times 2$ identity matrix and  the dimensionless damping constants are defined as $\alpha_1 \equiv \eta/r_c^2$, $\alpha_2 \equiv \eta \, \tilde{\alpha}_{\rm R}/r_c$, and $\alpha_3 \equiv \eta \, \tilde{\alpha}_{\rm R}^2$,  in analogy with the domain wall case where the core radius $r_c$ plays the role here as the characteristic length scale.  The coefficients $D_i$ depend on the core profile and are given by
\begin{align}
D_0 &= \pi \int_{0}^{\infty} dr \, \left(  r  \left(\partial_r \theta_0\right)^2 + \frac{\sin^2\theta_0}{r} \right), \\
D_1 &= 2\pi r_c^2 \int_{0}^{\infty} dr \, \frac{1}{r} \left(\partial_r \theta_0\right)^2 \sin^2\theta_0, \\
D_2 &= 2 \pi r_c  \int_{0}^{\infty} dr \, \frac{1}{r} \left(\partial_r\theta_0\right) \sin\theta_0 \left(  r \left(\partial_r\theta_0\right) \cos\theta_0   + \sin\theta_0  \right), \\
D_3 &= \pi \int_{0}^{\infty} dr \, \left( r \left(\partial_r\theta_0\right)^2  \cos^2\theta_0  +  \frac{\sin^2\theta_0}{r} \right),
\end{align}
where the expression for $D_0$ is a known result but $D_1, D_2$ and $D_3$ are new terms that arise from the nonlinear anisotropic damping due to RSOC.

The coefficients $a_{11}$ and $a_{22}$ are configuration-dependent and represent the chiral component of the Rashba-induced damping. For vortex-type spin textures ($c=1,3$ and $q=1$), $a_{11} = a_{22} = 0$, which indicates that the $\alpha_2$ term plays no role for such configurations. This is consistent with the result for Bloch domain walls discussed previously, since the vortex-type texture [Fig.~\ref{fig:vortex_config}(b)], particularly the vortex-type skyrmion [Fig.~\ref{fig:vortex_config}(c)], can be thought of as being analogous to a spin structure generated by a $2\pi$ revolution of a  Bloch wall about an axis perpendicular to the film plane. The rigid core approximation implies that fluctuations about the ground state are neglected, which is akin to setting  $\delta\phi(t) = 0$ in Eq.~(\ref{eq:dissfn_Bloch_lin}). As such, no contribution from $\alpha_2$ is expected for vortex-type textures. On the other hand, a finite contribution appears for hedgehog-type vortices and skyrmions ($q=1$), where $a_{11} = a_{22} = 1$ for $c=0$ and $a_{11} = a_{22} = -1$ for $c=2$. This can be understood with the same argument by noting that hedgehog-type textures can be generated by  revolving N\'eel-type domain walls. A summary of these coefficients is given in Table~\ref{tab:dyadic_coeffs}.
\begin{table}
\caption{Coefficients $a_{11}$ and $a_{22}$ of the chiral damping term in Eq.~(\ref{eq:dyadic}) for different vortex/skyrmion charges $q$ and chirality $c$.}
\label{tab:dyadic_coeffs}
\begin{ruledtabular}
\begin{tabular}{c|cccc|cccc}
  & \multicolumn{4}{ c| }{$q = 1$} & \multicolumn{4}{ c}{$q = -1$} \\
 $c$ & 0 & 1 & 2 & 3 & 0 & 1 & 2 & 3 \\
$a_{11}$ & 1 & 0 & $-1$ & 0 & $-1$ & $-1$ & 1 & 1 \\
$a_{22}$ & 1 & 0 & $-1$ & 0 & 1 & $-1$ & $-1$ & 1
\end{tabular}
\end{ruledtabular}
\end{table}

For antivortices ($q = -1$), it is found that the coefficients $a_{ii}$ are nonzero for all winding numbers considered. We can  understand this qualitatively by examining how the magnetization varies across the core along two orthogonal directions. For example, for $c=0$, the variation along the $x$ and $y$ axes across the core are akin to two N{\'e}el-type walls of different chiralities, which results in nonvanishing contributions to $a_{11}$ and $a_{22}$ but with opposite sign. The sign of these coefficients depends on how these axes are oriented in the film plane, as witnessed by the different chiralities $c$ in Fig.~\ref{fig:vortex_config}. Such damping dynamics is therefore strongly anisotropic, which may have interesting consequences on the rotational motion of vortex-antivortex dipoles, for example, where the antivortex configuration oscillates between the different $c$ values in time~\cite{Finocchio:2008kv}.

For vortex structures, we can provide numerical estimates of the different damping contributions $\alpha_i D_i$ by using the Usov ansatz for the vortex core magnetization,
\begin{equation}
\cos\theta_0 = %
\begin{dcases}
\frac{r_c^2-r^2}{r_c^2+r^2} & r \leq r_c \\
0 & r > r_c
\end{dcases}.
\end{equation}
Let $L$ represent the lateral system size. The coefficients $D_i$ are then found to be $D_0 = \pi \left[2 + \ln\left(L/r_c\right) \right]$, $D_1 = D_2 = 14\pi/3$, and $D_3 = \pi \left[4/3 + \ln\left(L/r_c\right) \right]$. We note that for $D_0$ and $D_3$, the system size $L$ and core radius $r_c$ appear as cutoffs for the divergent $1/r$ term in the integral. By assuming  parameters of $\alpha_0 = 0.1$, $\eta = 0.05$ nm$^2$, and $\alpha_R = 0.1$ eV nm, along with typical scales of $r_c = 10$ nm and $L = 1$ $\mu$m, the damping terms can be evaluated numerically to be $\alpha_0 D_0 \approx 2.1$, $\alpha_1 D_1 \approx 0.0073$,  $\alpha_2 D_2 \approx 0.19$, and $\alpha_3 D_3 \approx 6.4$. As for the domain walls, the Rashba term $\alpha_3 D_3$ is the dominant contribution and is of the same order of magnitude as the linear damping term, while the chiral term $\alpha_2 D_2$ is an order of magnitude smaller and the nonlinear term $\alpha_1 D_1$ is negligible in comparison.

For skyrmion configurations, a similar \emph{ansatz} can be used for the core magnetization,
\begin{equation}
\cos\left(\frac{\theta_0}{2}\right) = %
\begin{dcases}
\frac{r_c^2-r^2}{r_c^2+r^2} & r \leq r_c \\
0 & r > r_c
\end{dcases}.
\label{eq:skyrmion_core}
\end{equation}
We note that this differs from the ``linear'' profiles discussed elsewhere~\cite{Kiselev:2011cm}, but the numerical differences are small and do not influence the qualitative features of the dynamics. The advantage of the \emph{ansatz} in Eq.~(\ref{eq:skyrmion_core}) is that the integrals for $D_i$ have simple analytical expressions. Because  spatial variations in the magnetization for a skyrmion are localized only to the core, in contrast to the circulating in-plane moments of vortices that extend across the entire system, the damping constants $D_i$ have no explicit dependence on the system size. By using Eq.~(\ref{eq:skyrmion_core}), we find $D_0 = D_3 = 16\pi/3$, $D_1 = 496\pi/15$, and $D_2 = 52\pi/5$. By using the same values of $\alpha_0$, $\eta$, and $\alpha_\textrm{R}$ as for the vortices in the preceding paragraph, we find $\alpha_0 D_0 \approx 1.7$, $\alpha_1 D_1 \approx 0.052$, $\alpha_2 D_2 \approx 0.43$, and $\alpha_3 D_3 \approx 3.3$.

\section{\label{sec:discussion}Discussion and concluding remarks}

A clear consequence of the nonlinear anisotropic damping introduced in Eq.~(\ref{eq:nad}) is that it provides a mechanism by which the overall damping constant, as extracted from domain wall experiments, for example, can differ from the value obtained using linear response methods such as ferromagnetic resonance~\cite{Weindler:2014et}. However, the Rashba term can also affect the ferromagnetic linewidth in a nontrivial way. To see this, we consider the effect of the damping by evaluating the dissipation function associated with a spin wave propagating in the plane of a perpendicularly magnetized system with an amplitude $b(t)$ and wave vector $\mathbf{k}_{||}$. The spin wave can be expressed as  $\mathbf{m} = \left[ b(t) \cos(\mathbf{k}_{||}\cdot \mathbf{r}_{||}), b(t) \sin(\mathbf{k}_{||}\cdot \mathbf{r}_{||}), 1 \right]$, which results in a dissipation function per unit volume of
\begin{equation}
W_{\rm sw} = \frac{M_s}{2 \gamma} \dot{b}(t)^2 \left(  \alpha_0 + \alpha_3  +  \eta \, b(t)^2 \|  \mathbf{k}_{||}  \|^2  \right),
\end{equation}
where a term proportional to the chiral part $\eta \tilde{\alpha}_{\rm R}$ spatially averages out to zero. The Rashba contribution $\alpha_3 \equiv \eta \tilde{\alpha}_{\rm R}^2$ leads to an overall increase in the damping for linear excitations and plays the same role as the usual Gilbert term $\alpha_0$ in this approximation, which allows us to assimilate the two terms as an effective FMR damping constant, $\alpha_{\rm FMR} \approx \alpha_0 + \alpha_3$. On the other hand, the nonlinear feedback term proportional to $\eta$ is only important for large spin wave amplitudes and depends quadratically on the wave vector. This is consistent with recent experiments on permalloy films (in the absence of RSOC) in which the linear Gilbert damping was recovered in ferromagnetic resonance while nonlinear contributions were only seen for domain wall motion~\cite{Weindler:2014et}. This result also suggests that the large damping constant in ultrathin Pt/Co/Al$_2$O$_3$ films as determined by similar time-resolved magneto-optical microscopy experiments, where it is  found that $\alpha_{\rm FMR} = 0.1$--$0.3$~\cite{Schellekens:2013it}, may partly be due to the RSOC mechanism described here (although dissipation resulting from spin pumping into the platinum underlayer is also likely to be important~\cite{Beaujour:2006gz}). Incidentally, the nonlinear term $\eta \, b(t)^2$ may provide a physical basis for the phenomenological nonlinear damping model proposed in the context of spin-torque nano-oscillators~\cite{Tiberkevich:2007it}.

For vortices and skyrmions, the increase in the overall damping due to the Rashba term $\alpha_3$ can have important consequences for their dynamics. The gyrotropic response to any force, as described by the Thiele equation in Eq.~(\ref{eq:Thiele}), depends on the overall strength of the damping term. This response can be characterized by a deflection angle, $\theta_H$, that describes the degree to which the resulting displacement is noncollinear with an applied force. This is analogous to a Hall effect. By neglecting the chiral term $\alpha_2 D_2$, the deflection or Hall angle can be deduced from Eq.~(\ref{eq:Thiele}) to be
\begin{equation}
\tan{\theta_H} = \frac{G_0}{ \alpha_0 D_0 + \alpha_1 D_1 + \alpha_3 D_3 }, 
\label{eq:hallangle}
\end{equation}
where $G_0 =2 \pi$ for vortices and $G_0 =4 \pi$ for skyrmions. Consider the skyrmion profile and the magnetic parameters discussed in Section~\ref{sec:vorsky}. With only the linear Gilbert damping term ($\alpha_0 D_0$) the Hall angle is found to be $\theta_H = 82.3^{\circ}$, which underlies the largely gyrotropic nature of the dynamics. If the full nonlinear damping is taken into account [Eq.~(\ref{eq:hallangle})], we find $\theta_H = 68.3^{\circ}$, which represents a significant reduction in the Hall effect and a greater Newtonian response to an applied force. Aside from a quantitative increase in the overall damping, the presence of the nonlinear terms can therefore affect the dynamics qualitatively. Such considerations may be important for interpreting current-driven skyrmion dynamics in racetrack geometries, where the interplay between edge repulsion and spin torques is crucial for determining skyrmion trajectories~\cite{Fert:2013fq, Sampaio:2013kn}.

Finally, we conclude by commenting on the relevance of the chiral-dependent component of the damping term, $\alpha_2$. It has been shown theoretically that the Rashba spin-orbit coupling leading to Eq.~(\ref{eq:nad}) also gives rise to an effective chiral interaction of the Dzyaloshinskii-Moriya form~\cite{Kim:2013gm}. This interaction is equivalent to the interface-driven form considered earlier, which favors monochiral N{\'e}el wall structures in ultrathin films with perpendicular magnetic anisotropy. Within this picture, a sufficiently strong Rashba interaction should only favor domain wall or skyrmion spin textures with one given chirality as determined by the induced Dzyaloshinskii-Moriya interaction. So while some non-negligible differences in the chiral damping between vortices and skyrmions of different chiralities were found, probing the dynamics of solitons with distinct chiralities may be very difficult to achieve experimentally in material systems of interest.

\begin{acknowledgments}
The author acknowledges fruitful discussions with P. Borys, J.-Y. Chauleau, and F. Garcia-Sanchez. This work was partially supported by the Agence Nationale de la Recherche under Contract No. ANR-11-BS10-003 (NanoSWITI) and No. ANR-14-CE26-0012 (ULTRASKY). 
\end{acknowledgments}

\bibliography{articles}

\end{document}